\documentclass[twocolumn,reprint,amsmath,amssymb,aip]{revtex4-1}

\usepackage{graphicx}
\usepackage{dcolumn}
\usepackage{bm}
\usepackage{braket}
\usepackage{color}
\usepackage{mathtools}

\begin{document}

\title{Controlled creation of three-dimensional vortex structures in Bose--Einstein condensates using artificial magnetic fields}

\author{James Schloss}
\affiliation{OIST Graduate University, 904-0495 Okinawa, Japan}

\author{Peter Barnett}
\affiliation{OIST Graduate University, 904-0495 Okinawa, Japan}
\affiliation{Department of Physics, University of Otago, Dunedin 9054,New Zealand}

\author{Rashi Sachdeva}
\affiliation{OIST Graduate University, 904-0495 Okinawa, Japan}

\author{Thomas Busch}
\affiliation{OIST Graduate University, 904-0495 Okinawa, Japan}

\begin{abstract}
The physics of quantized vortex excitations in atomic Bose-Einstein condensates has been extensively studied  in recent years.
Although simple vortex lines are relatively easy to create, control, and measure in experiments, it is a lot more difficult to do the same for vortex ring structures.
Here we suggest and explore a method for generating and controlling superfluid vortex rings, vortex ring lattices, and other three dimensional vortex structures in toroidally-trapped superfluid Bose--Einstein condensates by using the artificial magnetic field produced by an optical nanofiber.
The presence of the fiber also necessitates a multiply-connected geometry and we show that in this situation the presence of these vortex structures can be deduced from exciting the scissors mode of the condensate.
\end{abstract}


                              
\maketitle

\section{Introduction}

Atomic Bose--Einstein condensates (BECs) are superfluids consisting of neutral, bosonic atoms that have been cooled and condensed into the macroscopic ground state of an external potential~\cite{pethick_smith_2008}.
They have been shown to support a large number of flow-related excitations, with the most common ones being quantized vortex lines and vortex rings~\cite{madison2000,abo2001, wacks2014, anderson2001, bulgac2014, ku2016, matthews1999, yefsah2013}.
However, vortices with higher winding numbers are unstable in singly-connected condensates, which means that increasing the amount of angular momentum imparted on the superfluid will lead to an increasing number of vortices with a winding number of one.
These vortex lines interact repulsively and larger numbers will eventually arrange themselves in the form of a triangular, Abrikosov lattice \cite{abo2001}, similar to the behaviour known for Type II superconductors \cite{abrikosov1957}.
Due to the quantization and the homogeneity in winding numbers, single component condensates are  often suggested and used for studying superfluid turbulence \cite{tsubota2014,barenghi2014,seo2017,navon2016}.

In a finite-sized atomic condensate without dissipative effects, all vortex lines must have a finite length and either start and end at the cloud surface \cite{madison2000} or reconnect onto themselves\cite{barenghi2014}.
Complex, three dimensional vortex topologies beyond vortex lines cannot be easily created by stirring or rotating a BEC because the vortex lines generated in this way must follow the axis of rotation; therefore, to consistently control and generate vortex rings or other topological structures, methods beyond stirring are required and only a small number of experimental realizations of these have been reported \cite{anderson2001,yefsah2013}.

In most cases, including in most theoretical proposals, vortex ring generation in BECs relies on dynamic processes that do not create eigenstates of the system.
These include using the decay of dark solitons in multicomponent condensates \cite{anderson2001} via the snake instability \cite{ruostekoski2001}, direct density engineering \cite{shomroni2009, ruostekoski2005}, or the collision of symmetric defects \cite{ginsberg2005}.
Other theoretical proposals have considered interfering two BECs \cite{jackson1999}, using spatially dependent Feschbach resonances \cite{pinsker2013}, or direct phase imprinting methods \cite{ruostekoski2001}.
It should be noted that for inhomogeneously trapped BECs, vortex ring structures are known to be unstable, which has led to difficulties in their experimental observation \cite{abad2008}.
In addition, the direct absoption imaging techniques employed in the field of BECs are not well suited to determine whether a three dimensional vortex structure is present in an experimental system or not.

Another method to induce rotational effects in a BEC is through the introduction of artificial magnetic fields, which can be created, for example, by the interaction between an atomic system in a dressed state  and an electric field that is tuned near an atomic resonance frequency \cite{dalibard2011}.
In this case, instead of following an axis of rotation, the vortices follow along the artificial magnetic field lines, which allows one to stably generate complex vortex structures by modulating the geometry of the magnetic field profiles.

In this work we will discuss a system that allows for the tunable creation of artificial magnetic fields based on electromagnetic fields that vary strongly over short distances.
Such behavior can be found in the near-field regime on the surface of a dielectric system, when light undergoes total internal reflection \cite{mochol2015}.
For the generation of vortex rings, a suitable dielectric system is the optical nanofiber, which is an optical element that has several propagation modes to allow for the configurable generation of evanescent fields.
Nanofiber systems can be created by heating and stretching optical fibers until their thinnest region is roughly hundreds of nanometers in diameter \cite{ward2006, tong2003}.
At this scale, the wavelength of light is larger than the diameter of the fiber and the strength of the evanescent field is significantly enhanced \cite{yariv1976}.
The form of the evanescent field varies significantly depending on the optical modes propagating through the nanofiber, and we will show that this can be used to generate interesting and tunable artificial magnetic fields.

\begin{figure}[t]
\begin{center}
\includegraphics[width=0.6\linewidth]{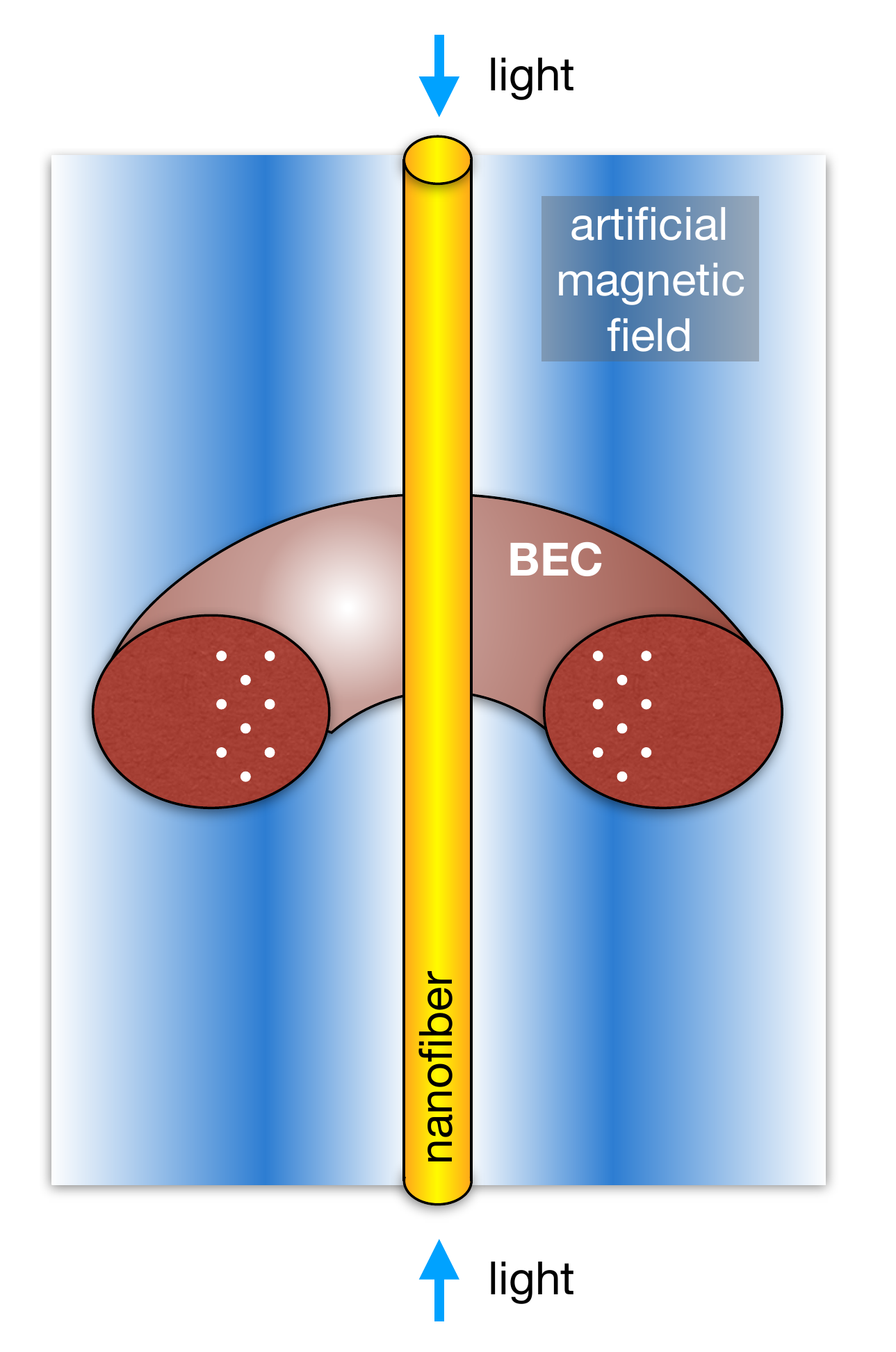}
\end{center}
\caption{Schematic of the system. Blue or red-detuned light is sent into the nanofiber (yellow), creating an evanescent field and artificial magnetic field (blue) that influences the BEC (maroon) held by a toroidal trapping potential. If the artificial magnetic field strength is greater than a threshold value, vortex rings (white) will appear and begin to arrange themselves into a triangular lattice.}
\label{Fig:device}
\end{figure}

Optical nanofibers are already used in many different experiments with ultracold atoms \cite{vetsch2010, lacro_te2012, nieddu2016, sague2007, russel2011, kumar2015}, and trapping potentials at around 200nm from the fiber surface can be created by using two differently detuned input fields \cite{kien2004, phelan2013}.
Our proposed setup will allow for the creation of vortex rings in BECs that are trapped toroidally around the fiber at roughly the same distance by coupling the BEC to the evanescent field created by different modes propagating through the nanofiber \cite{sachdeva2017}.
A schematic of system is depicted in Fig.~\ref{Fig:device}.

In addition, we also discuss a mechanism to detect vortex rings by exciting the scissors mode in an elliptic-toroidal system \cite{cozzini2003, guery1999, marago2000}.
This can be done by slightly tilting the trapping geometry radially from the center of the torus, which will cause the BEC to oscillate in and out due to the new potential.
Without a vortex present, this oscillation possesses a single frequency, whereas in the presence of a vortex ring it will contain two frequencies that average to the vortex-less oscillation frequency, similar to scissors mode oscillation frequencies in two dimensional, elliptically-trapped BECs \cite{smith2004, zambelli1998, stringari2001}.
While in simply-connected condensates, vortex rings cannot be detected by this method, we show that in our multiply-connected, elliptic-toroidal geometry, this mode allows for the detection of three dimensional vortex structures.

The manuscript is organized as follows: In Section~\ref{sec:bg}, we will discuss how BECs interact with the evanescent field profiles generated by the optical nanofiber.
Then, in Section~\ref{sec:vortex}, we will show simulated results of the vortex configurations that can be generated and discuss 
in Section~\ref{sec:dynamics} how to detect whether a vortex ring exists by exciting the scissors mode in an elliptic-toroidal geometry.
Finally, in Section~\ref{sec:discussion}, we will discuss potential extensions of the suggested system.

\section{Bose--Einstein condensates in the presence of an optical nanofiber}
\label{sec:bg}

The superfluid properties of atomic Bose-Einstein condensates are captured by the Gross-Pitaveskii Equation (GPE), which describes the evolution of the condensate wave-function in the mean field limit as \cite{pethick_smith_2008}
\begin{equation}
	i\hbar \frac{\partial \Psi}{\partial t} = \left[\frac{(p-m\mathbf{A}({\bf r}))^2}{2m} + V_{\text{trap}}({\bf r}) + g|\Psi|^2 \right] \Psi.
\end{equation}
Here $p = -i\hbar\frac{\partial}{\partial \textbf{r}}$ is the standard momentum operator and the kinetic energy term also accounts for the presence of a spatially inhomogeneous artificial vector potential, $\mathbf{A(r)}$.
The potential term  $V_{\text{trap}}({\bf r})$ describes an external trap and the non-linear term accounts for the scattering interaction between the atoms.
Its strength is given by $g = \frac{4\pi\hbar^2a_s}{m}$, where $a_s$ is the scattering length of the atomic species and $m$ its mass.
The artificial vector potential can take many forms, and for our purpose we choose a description in terms of Berry's connection \cite{dalibard2011}
\begin{equation}
  \mathbf{A} = i\hbar \braket{\Psi_l | \nabla \Psi_l},
\end{equation}
where $\Psi_l$ is the atomic wavefunction in some dressed state $l$.
Since we will be considering two-state atoms in the presence of an optical field,
the relevant dressed states can be written within the rotating wave approximation as \cite{mochol2015}
\begin{align}
\ket{\Psi_1({\bf r})} &= 
\begin{pmatrix*}[l] 
    \cos[\Phi({\bf r})/2] \\
     \sin[\Phi({\bf r}) /2]e^{i\phi(z)} 
\end{pmatrix*},\\
\ket{\Psi_2({\bf r})} &= 
\begin{pmatrix*}[l]
    -\sin[\Phi({\bf r}) /2]e^{-i\phi(z)} \\
    \quad\cos[\Phi({\bf r})/2]
\end{pmatrix*},
\end{align}
where $\phi(z)$ is the phase of the optical field and $\Phi({\bf r}) = \arctan(|\kappa({\bf r})|/\Delta)$, with $\Delta = \omega_0 - \omega$ being the detuning and $\kappa({\bf r}) = \mathbf{d} \cdot \mathbf{E}({\bf r}) / \hbar$ being the Rabi frequency.
The atomic dipole moment is given by $\mathbf{d}$ and $\mathbf{E(\mathbf{r})}$ is the electric field.

The form of the artificial vector potential, $\mathbf{A(r)}$, is therefore determined by the form of the optical fields, and for the nanofiber system it can be controlled by choosing specific optical modes to travel through the fiber.
The artificial magnetic field associated with the spatially varying  artificial vector potential is then given by  $\mathbf{B}= \nabla \times \mathbf{A}$, and when the magnetic field lines penetrate the condensates, an artificial Lorentz force will lead to the creation of vortices around these field lines. 

To determine which modes will propagate in an optical fiber, one needs to calculate the $V$-number, which is given by $V = k_0a\sqrt{n_1^2 - n_2^2}$.
Here $a$ is the fiber radius, $n_1$ is the refractive index of the fiber, $n_2$ is the refractive index of the cladding, and $k_0 = \omega/c$ with $\omega$ being the frequency of the input light beam.
In this case, the fiber has been tapered such that the cladding has become the vacuum with $n_2=1$.
Higher order modes can only be sustained if $V > V_c \simeq 2.405$, and below this value only the fundamental HE$_{11}$ mode can propagate.
The $V$-number can easily be controlled by choosing the fiber radius \cite{nieddu2016, kumar2015}

Using cylindrical coordinates, the evanescent field around the nanofiber corresponding to the HE$_{\ell m}$ mode with circular polarization is given by \cite{minogin2010}
\begin{align}
	E_r &= iC[(1-s)K_{\ell-1}(qr) + (1+s)K_{\ell+1}(qr)]e^{i(\omega t- \beta z)}, \\
	E_\phi &= -C[(1-s)K_{\ell-1}(qr) - (1+s)K_{\ell+1}(qr)]e^{i(\omega t- \beta z)}, \\
	E_z &= 2C(q/\beta)K_\ell(qr)e^{i(\omega t - \beta z)},
\end{align}
where
\begin{align}
	s &= \frac{1/h^2a^2 + 1/q^2a^2}{J_\ell'(ha)/[haJ_\ell(ha)]+K_\ell'(qa)/[qaK_\ell(qa)]}, \\
	C &= \frac{\beta}{2q}\frac{J_\ell(ha)/K_\ell(qa)}{\sqrt{2\pi a^2(n_1^2N_1+n_2^2N_2)}},
\end{align}
and 
\begin{align}
N_1 =&\frac{\beta^2}{4h^2}\Big[(1-s)^2\left[J_{\ell-1}^2(ha)+J_\ell^2(ha)\right]\nonumber \\
   & \qquad+(1+s)^2\left[J_{\ell+1}^2(ha)-J_\ell(ha)J_{\ell+2}(ha)\right]\Big] \nonumber\\
   & +\frac12\left[J_\ell^2(ha) - J_{\ell-1}(ha)J_{\ell+1}(ha)\right], \\
N_2 =& \frac{J_\ell^2(ha)}{2K_\ell^2(qa)}\Bigg( \frac{\beta^2}{4q^2}\Big[(1-s)^2\left[K_{\ell-1}^2(qa)-K_\ell^2(qa)\right] \nonumber\\
   & \qquad\qquad\quad-(1+s)^2\left[K_{\ell+1}^2(qa)-K_\ell(qa)K_{\ell+2}(qa)\right]\Big] \nonumber\\
   & \qquad\qquad -\frac12\left[K_\ell^2(qa) + K_{\ell-1}(qa)K_{\ell+1}(qa)\right]\Bigg ).
\end{align}
The mode geometry is given by $J_n(x)$, the Bessel function of the first kind, $K_n(x)$, the modified Bessel function of the second kind, and $\beta$, the propagation constant of the fiber.
The scaling factors are given by $q = \sqrt{\beta^2-n_2^2k_0^2}$ and $h = \sqrt{n_1^2k_0^2 - \beta^2}$, the normalisation constant is $C$ and $s$ is a dimensionless parameter.

\begin{figure*}[tb]
 \includegraphics[width=0.8\linewidth]{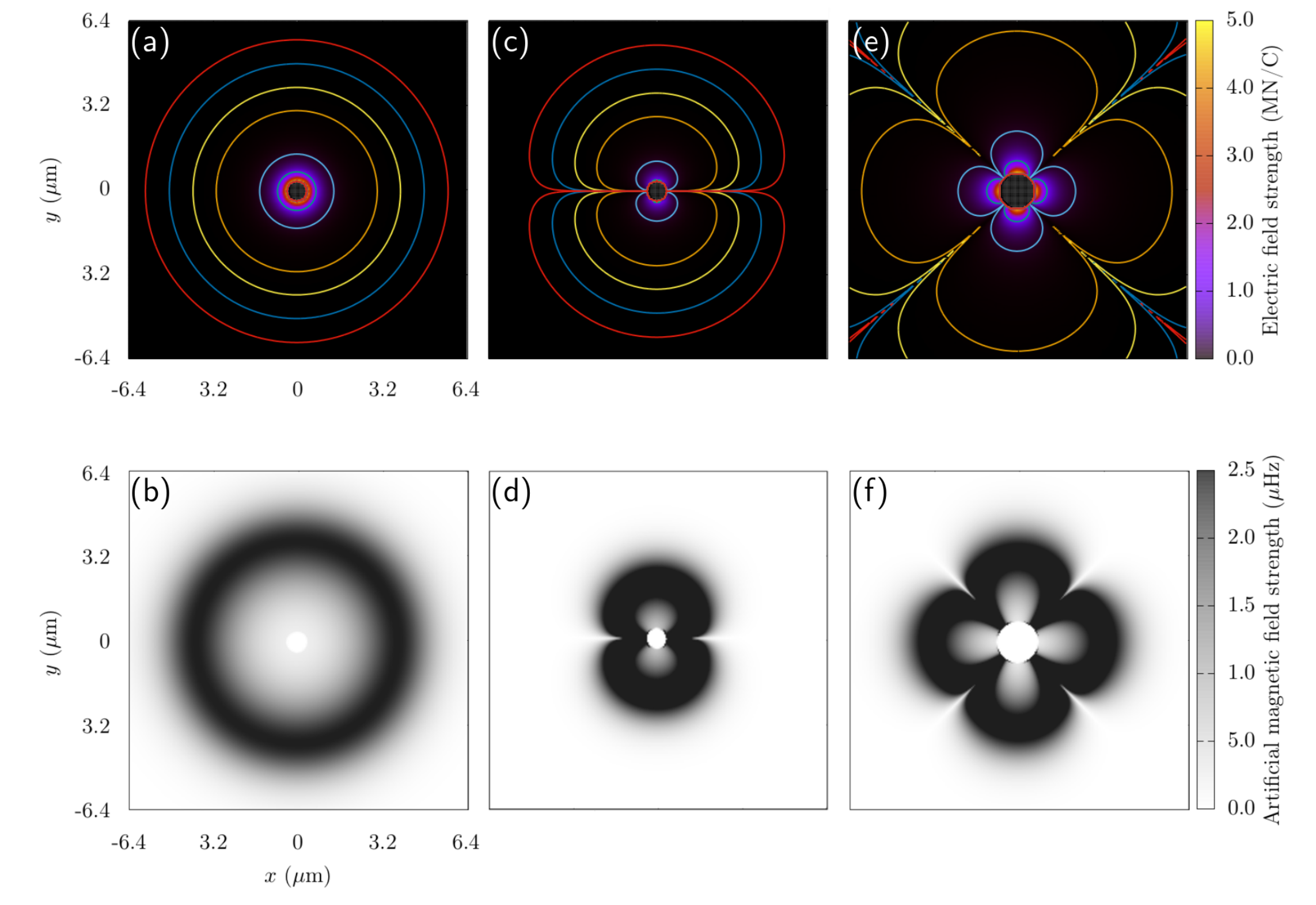}
	 \caption{Images of electric and artificial magnetic field profiles for [(a) and (b)] the fundamental HE$_{11}$ mode with circular polarization, [(c) and (d)] the HE$_{11}$ mode with linear polarization, and [(e) and (f)] the HE$_{21}$ mode with linear polarization. For these calculations, the input power is 372~nW in (a) and (b) , 16~nW in (c) and (d), and 418~nW in (e) and (f). For the HE$_{11}$ mode, the nanofiber radius is 200~nm with blue-detuned light of 700~nm, and for the HE$_{21}$ mode, the nanofiber radius is 400~nm with red-detuned light of 980~nm}
 \label{fig:EtoB}
\end{figure*}

When the input light field is linearly polarized, it is convenient to write the  cartesian components of the evanescent electric field as 
\begin{align}
E_x =& \sqrt 2 C\Big[(1-s)K_{\ell-1}(qr)\cos(\phi_0)\nonumber \\
      &\qquad+(1+s)K_{\ell+1}(qr)\cos(2\phi-\phi_0)\Big]e^{i(\omega t - \beta z)}, \\
E_y = &\sqrt 2 C\Big[(1-s)K_{\ell-1}(qr)\sin(\phi_0)\nonumber \\
      &\qquad+(1+s)K_{\ell+1}(qr)\sin(2\phi-\phi_0)\Big]e^{i(\omega t - \beta z)}, \\
E_z = & 2\sqrt 2 i C(q/\beta)K_\ell(qr)\cos(\phi - \phi_0)e^{i(\omega t - \beta z)}.
\end{align}
Here $\phi_0$ determines the orientation of polarization, with $\phi_0 = 0$ being along the $x$ axis and $\pi/2$ being along the $y$ axis.
The artificial vector potential produced by such evanescent fields around an optical nanofiber is then given by \cite{sachdeva2017}
\begin{equation}
\mathbf{A} = \hat{z} \hbar \kappa_0 (n_1 + 1) \tilde{s} \left[\frac{|d_rE_r + d_{\phi}E_{\phi} + d_zE_z|^2}{1 + \tilde s^2|d_rE_r + d_{\phi}E_{\phi} + d_zE_z|^2} \right],
\end{equation}
where $\tilde s = \frac{|\mathbf{d}\cdot\mathbf{E}|}{\hbar |\Delta|}$ and
the corresponding magnetic field $\mathbf{B} = \nabla \times \mathbf{A}$ can be calculated to be
\begin{align}
\mathbf{B} =& \frac{\hbar \kappa_0 s^2(n_1 + 1)}{(1+\tilde s^2|d_rE_r + d_{\phi}E_{\phi} + d_zE_z|^2)^2} \nonumber\\
&\times \bigg[ \hat\phi  \frac{\partial}{\partial r} |d_rE_r + d_{\phi}E_{\phi} + d_zE_z|^2 \nonumber\\
&\qquad- \hat r \frac{1}{r} \frac{\partial}{\partial \phi} |d_rE_r + d_{\phi}E_{\phi} + d_zE_z|^2 \bigg ].
\end{align}
This shows that the $\mathbf{B}$ field has only components in the $\hat \phi$ and $\hat r$ directions, which means that all field lines lie in the horizontal plane if the fiber is aligned along the vertical $\hat z$ direction.

\begin{figure*}[tb]
\includegraphics[width=\textwidth]{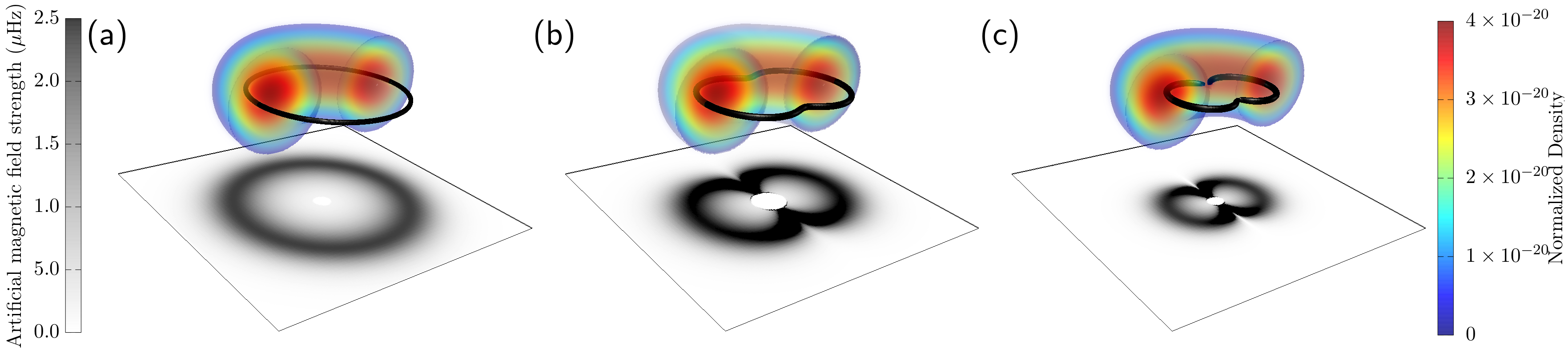}
\caption{Vortex configurations for different magnetic field profiles from the nanofiber for the fundamental HE$_{11}$ mode with (a) circular polarization, (b) elliptical polarization, and (c) linear polarization along the $\hat y$ direction. The vortex distributions have been found via an isosurface on the Sobel filtered wavefunction density for a $^{87}$Rb BEC and all optical fiber fields are normalized and for a nanofiber of 200~nm in radius with blue-detuned light of 700nm. The magnetic field profiles shown in the shaded region beneath wavefunction density are similar to those in Figure~\ref{fig:EtoB}(b) and (d).}
\label{fig:VortexRings}
\end{figure*}

For a BEC that is trapped cylindrically around a nanofiber one can therefore expect to find vortex structures that wrap around the nanofiber and potentially close in on themselves in the form of vortex rings; however, depending on the exact form of the evanescent mode other structures are possible as well.
Modulating the value of $\tilde s$ allows one to change the amplitude and range of the magnetic field and thereby change the size and shape of the generated vortex structures \cite{sachdeva2017}.
In the following we will focus on three different evanescent field configurations: the fundamental HE$_{11}$ mode with circular polarization, the HE$_{11}$ mode with linear polarization, and the HE$_{21}$ mode with linear polarization.
The electric field configurations and their corresponding artificial magnetic fields can be seen in Fig.~\ref{fig:EtoB}.
It is notable that using the circularly polarized fundamental HE$_{11}$ mode leads to cylindrically symmetric electric (Fig.~\ref{fig:EtoB}(a)) and artificial magnetic field configurations (Fig.~\ref{fig:EtoB}(b)), whereas using linearly polarized light leads to a lobed structure for both quantities (see Figs.~\ref{fig:EtoB}(c) and (d)).
When using linearly-polarized light with the higher-order HE$_{21}$ mode, an even more complex structure composed of four petals appears (see Figs.~\ref{fig:EtoB}(e) and (f)) and the broken rotational symmetry suggests these fields will lead to the appearance of non-standard flow excitations. 
While using even higher order modes or interfering different modes can lead to even more complicated fields  \cite{sachdeva2017}, we concentrate here on the three examples above, as they demonstrate the large range of fundamental possibilities the system allows for.

\begin{figure}[tb]
 \includegraphics[width=\linewidth]{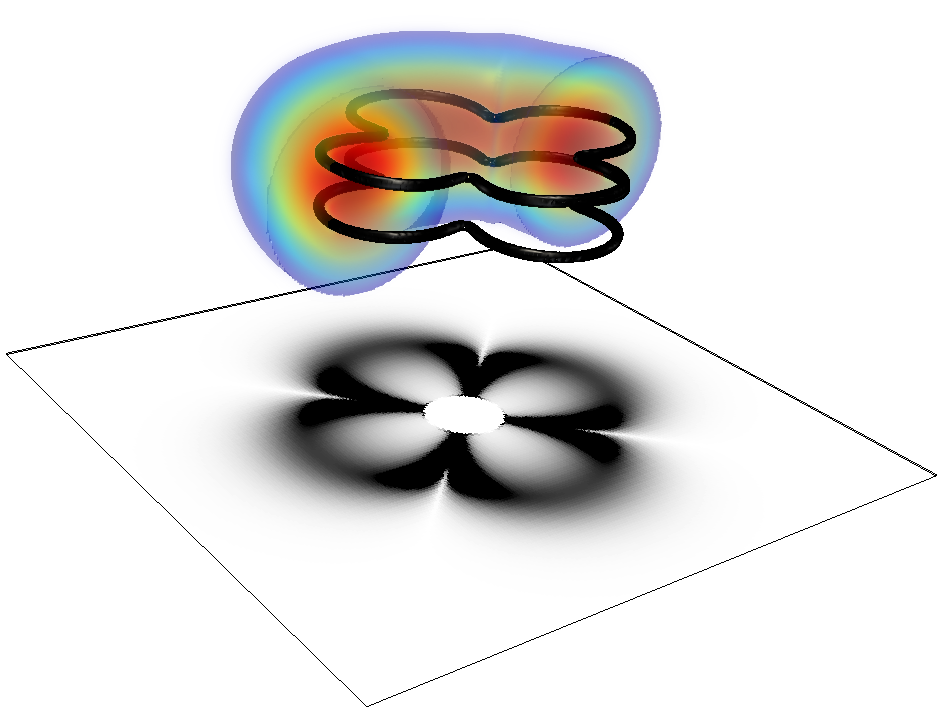}
 
 \caption{Vortex configuration for the HE$_{21}$ mode with linear polarization along the $\hat y$ direction. The vortex distributions have been found via an isosurface on the Sobel filtered wavefunction density for a $^{87}$Rb BEC and the optical fiber fields are for a nanofiber of 400~nm in diameter for the bottom image with red-detuned light of 980~nm. The magnetic field profile is similar to the one shown in Figure~\ref{fig:EtoB}(f), and has been calculated for a nanofiber of 400~nm in radius with red-detuned light of 980~nm.}
 \label{fig:HE21_3d}
\end{figure}

\begin{figure}
\hspace{-0.725cm} \includegraphics[width=0.8\linewidth]{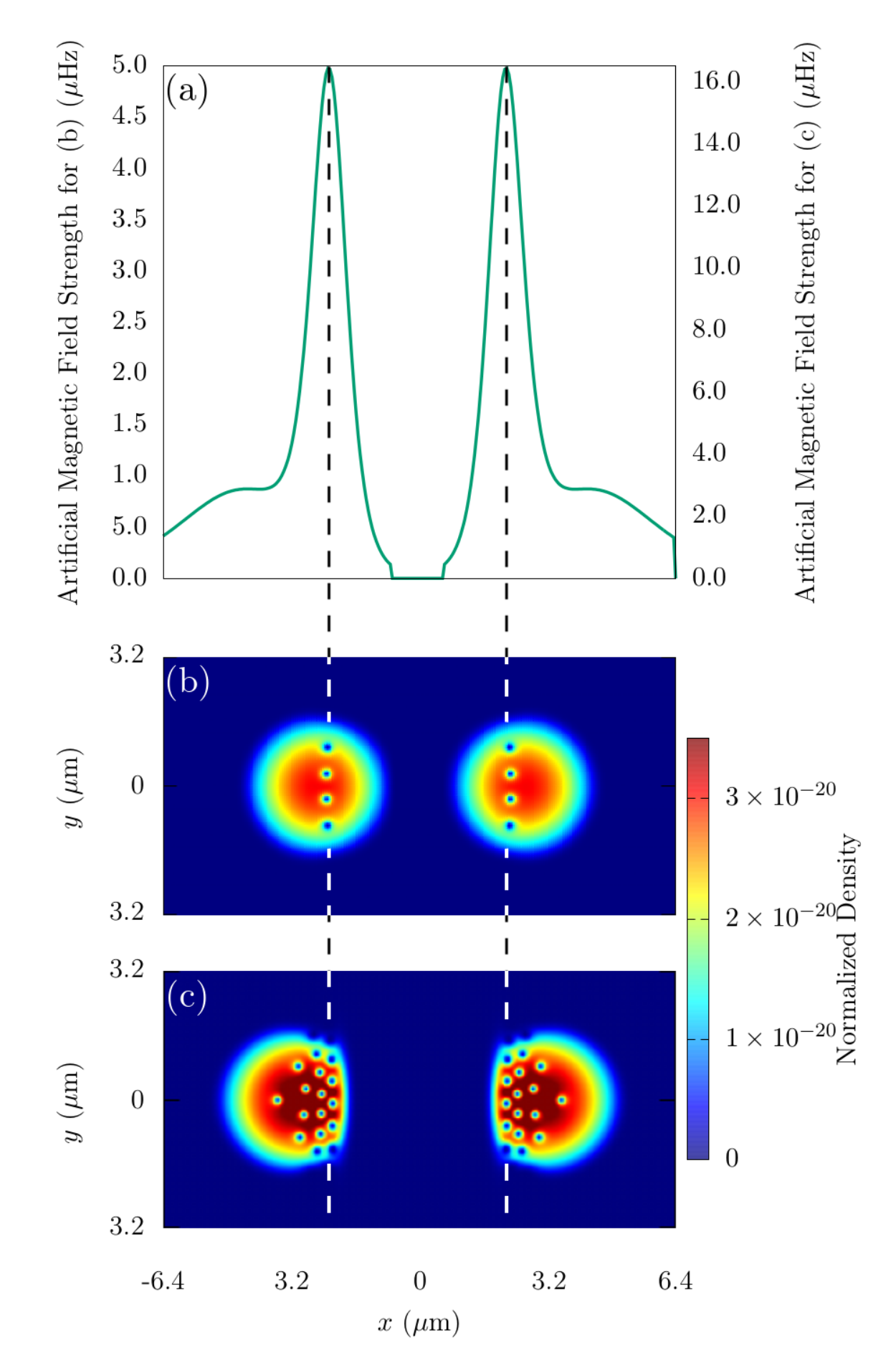}
\caption{(a) The magnetic field profile along the $x$-direction for the fundamental HE$_{11}$ mode with circular polarization outside a fibre of 200~nm radius. Note that for this mode and polarisation the whole system is azimuthally symmetric. For weak fields (see (b)) this leads to a small number of vortices that align along the line at which the magnetic field is maximal and for larger fields (see (c)) more vortex rings appear that form the beginning of an Abrikosov lattice.
The optical fiber field and wavefunction density have been normalized and are for a nanofiber of 200~nm in diameter with blue-detuned light of 700nm and a $^{87}$Rb BEC respectively.}
\label{fig:field_triangular}
\end{figure}

\section{Vortex configurations}
\label{sec:vortex}

To determine the vortex states that can be created by the evanescent fields around a nanofiber we solve the full three-dimensional Gross-Pitaevskii equation for a condensate trapped toroidally around the fiber.
For this, we use the GPUE codebase \cite{gpue} to describe a $^{87}$Rb condensate with $1\times10^5$ atoms with a scattering length of $a_s=4.76 \times 10^{-9}$ m on a three-dimensional grid of $256^3$ points with a spatial resolution of 50 nm. 
To clearly highlight the effects of the artificial magnetic fields, we assume a generic, external toroidal trapping around the fiber given by 
\begin{equation}
V_\text{trap} = m(\omega_r^2(r-\eta)^2 + \omega_z^2z^2),
\label{eqn:potential}
\end{equation}
where we chose the trapping frequencies in the $r$ and $z$ directions to be $\omega_r = \omega_z = 7071$Hz to match typical experimental conditions in fiber trapping \cite{vetsch2010}. 
The parameter $\eta$ defines the distance of the center of the toroidal condensate from the center of the fiber and is chosen such that the atoms are trapped outside the reach of the van-der-Waals potential of the fiber.
For simulations of the HE$_{11}$ mode, we assume a fiber radius of 200\,nm and we use $\eta = 3.20$\,$\mu$m to create a toroidal BEC with an inner radius roughly 300\,nm from the fiber surface. 
To simulate the effects of higher-order HE$_{21}$ modes, we assume an increased fiber radius of 400 nm, with all other parameters remaining the same. 
This creates a toroidal BEC with an inner radius roughly 150 nm from the fiber surface.

As a first example, we study the fundamental HE$_{11}$ mode with circular polarization, which is perfectly azimuthally symmetric.
One can therefore expect to find vortex lines that wrap around the fiber at a constant radius and reconnect onto themselves.
This is confirmed in Fig.~\ref{fig:VortexRings}(a), where we show the equilibrium solution for a field strength that leads to exactly one vortex ring.
For linearly polarized HE$_{11}$ modes, the circular symmetry is broken and one can see from Fig.~\ref{fig:VortexRings}(c) that the vortex lines bend towards the inner edge of the condensate, creating two vortex lobes.
This can be easily understood by realizing that the vortex lines have to follow lines of constant magnetic fields, which in these areas also bend towards and vanish into the fiber surface.
However, this also means that the field lines come very close when approaching the surface and careful examination of the condensate density shows that the vortex lines do not follow the field lines into the fiber surface, but rather connect to the neighbouring lobe when they approach each other within a healing length.
In fact, one can continuously go from the circular to the linear setting by considering elliptically polarized light, which leads to vortex rings that are deformed and interpolate between the azimuthally symmetric and fully folded in structure (see Fig.~\ref{fig:VortexRings}(b)). 
Finally, for the linearly polarized HE$_{21}$ mode, the superfluid system responds by creating vortex lines arranged in a four-petal shape, again mimicking the geometry of the artificial magnetic field (see Fig.~\ref{fig:HE21_3d}(a)).
For this situation we also show it is possible to create multiple of these vortex structures by increasing the field strength and that for low densities of these structures, they arrange themselves on top of the maximum of the inhomogeneous B-field inside the condensate (see Fig.~\ref{fig:field_triangular}(b) for the HE$_{11}$ mode).

To study control of multiple vortex structures with this system, we show that by increasing the B-field strength even further for the HE$_{11}$ mode, we can create an even larger number of vortex rings, which at a certain density, make a transition to arranging themselves in a triangular geometry, forming the equivalent of the famous Abrikosov lattice (see Figs.~\ref{fig:field_triangular}(a), (b) and (c)). However, as the artificial magnetic field is strongly inhomogeneous, this lattices forms close to and around the maximum of the  magnetic field.

It is therefore clear that one can control the shape of each vortex structure and their number by purely controlling the optical fields that are fed into the fiber, and that artificial magnetic fields around optical fibers provide unprecedented control for the creation of vortex ring-like structures.
In fact, because all optical fields can also be time dependent, this system can potentially be used for studies of the dynamical properties of these rings; however, in the latter case, additional care needs to be taken as high magnetic field values change the potential geometry of the atoms in the BEC due to a coupling between the artificial vector potential $\mathbf{A}$, and the trapping potential, $V_\text{trap}$.
In this case, the external potential $V_\text{trap}$ gets modified by a term proportional to $\mathbf{A}^2$, which has an effect on the condensate density beyond exciting rotation.
Time-dependent changes to the $\mathbf{A}$ field through changes in the laser intensity therefore also lead to phonon excitations, which in turn have an influence on the vortex line dynamics.
However, studies of the response of the wavefunction density to time-dependent artificial magnetic fields go beyond the scope of this work.
Nevertheless, let us stress, that for constant optical fields these (deformed) vortex-ring structures are stable and unique to creating vortex rings with artificial magnetic fields.
They cannot be excited using simple rotation in singly connected potentials.

\section{Dynamic vortex detection and scissor modes}
\label{sec:dynamics}

\begin{figure}
\includegraphics[width=0.5\textwidth]{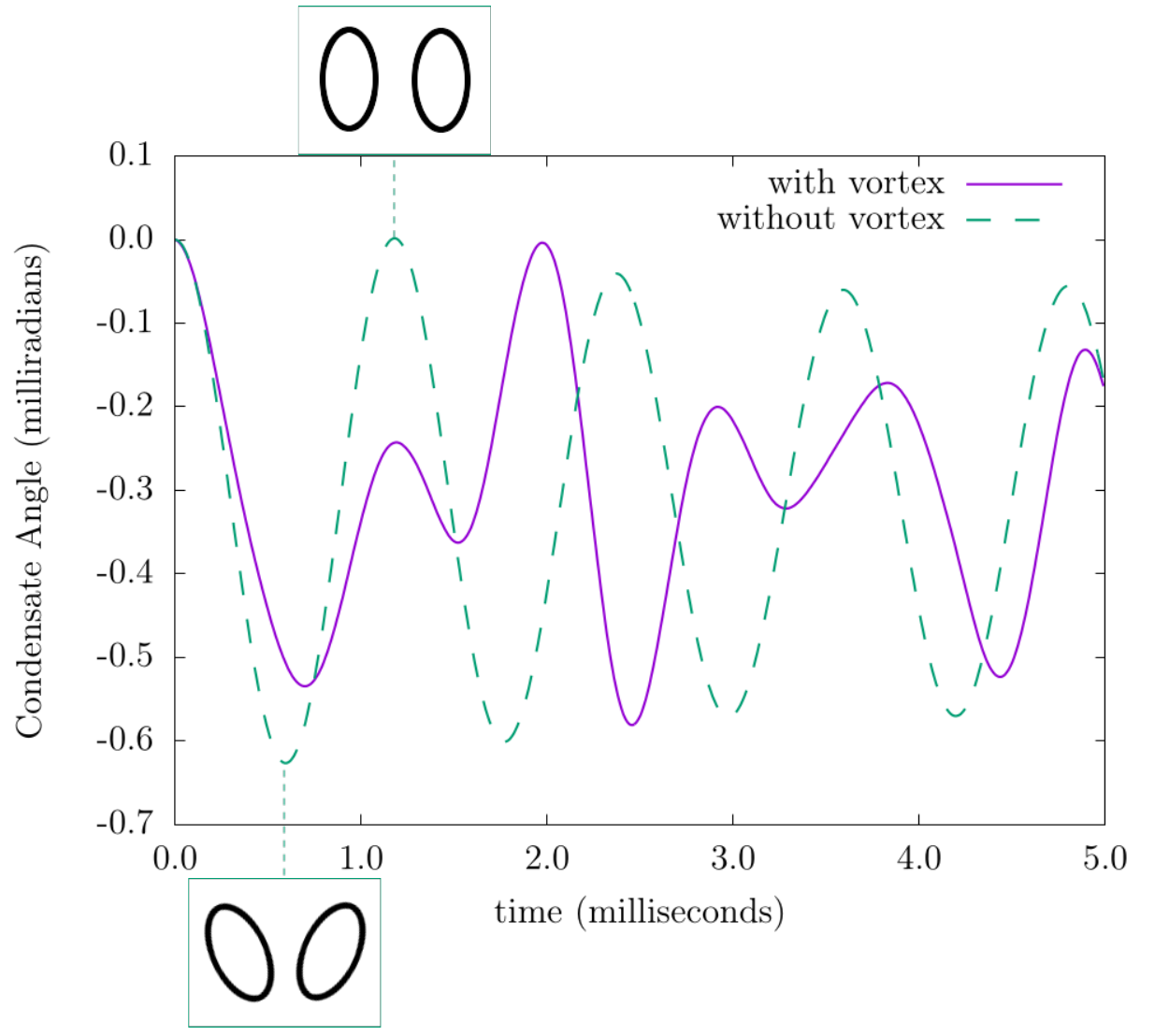}
\label{fig:scissors}
\caption{\label{fig:scissors} Angle of the condensate axis after excitation of the toroidal scissors mode for an elliptic-toroidal BEC with a single vortex ring (purple, solid) and without a vortex present (cyan, dashed). Depictions of two dimensional slices for the scissors mode without a vortex are shown in the insets. Here we see that the scissors mode causes oscillation in and out towards the center of the system, and that two distinct frequencies are present in the curve for the condensate carrying a vortex ring. }
\end{figure}

Observing the presence of vortex rings in a three dimensional BEC is a difficult problem, as absorption spectroscopy usually only provides a picture of an integrated two-dimensional density.
However, due to the unique geometry of the system we describe here, one can identify whether vortex rings are present by exciting the scissors mode of the condensate \cite{cozzini2003, guery1999, marago2000}. 
This works for systems trapped in elliptically toroidal geometries ($\omega_z < \omega_r$), which is a simple generalization of the above discussion. The scissors mode then gets excited by modifying the external potential with a rotation in the $rz$-plane
\begin{equation}
	V = V_{\text{trap}}(r, \theta, z) -m\omega_0^2\alpha rz,
\end{equation}
where $\alpha = 2\epsilon\theta$ is a coefficient related to the tilting angle, $\epsilon$ is the deformation of the trap in the $rz$ plane and $\theta$ is the angle at which the original trap was aligned at.
For a small initial angle of $\theta_0$ this change in the potential causes a BEC without rotation to oscillate back and forth in the trap with a frequency given by \cite{stringari2001}
\begin{equation}
	\omega_{\text{scissors}} = \sqrt{\omega_r^2+\omega_z^2}.
\end{equation}
If, however, this mode is excited for a BEC that contains a vortex line, the oscillation will be strongly influenced by the currents inside the condensate and two different frequencies ($\omega_+$ and $\omega_-$) appear in the oscillation spectrum \cite{smith2004, zambelli1998, stringari2001}.
The splitting of these two frequencies, when small compared to the frequency of the scissors mode without a vortex present,
can be written as \cite{zambelli1998}
\begin{equation}
\omega_{+} - \omega_{-} = \frac{\langle l_z \rangle}{m\langle r^2 + z^2 \rangle},
\label{eqn:splitting}
\end{equation}
where $\langle l_z \rangle$ is the average angular momentum per particle.
Calculating these values for our system for $\omega_r = 4242$Hz, and $\omega_z = 2828$Hz then leads to $\omega_{\text{scissors}} = 5090$Hz, $\omega_{-} = 3765$Hz, and $\omega_{+} = 6415$Hz, which are very close to the values observed in the numerical simulations shown in Fig.~\ref{fig:scissors}.
However, one can also see from this figure that the oscillation is not perfect and seems to decay over time.
This is due to the above mentioned modification of the trapping potential by the artificial vector potential, which leads to a deviation from the perfect elliptical toroidal shape used in the derivation of eq.~\eqref{eqn:splitting}.
Nevertheless, it does not effect the main argument. 

It is worth noting that this method cannot be used to detect a vortex ring inside a simply connected condensate, as in this situation the flow around the vortex line has no preferred direction.
However, in the toroidal shape each radial slice can be seen as a two-dimensional elliptical BEC with a single vortex, and the system will therefore exhibit the scissors mode frequency as expected.
While in principle the excitation of the scissors mode can also be used to detect Abrikosov vortex-ring lattice, the fact that the inhomogeneous artificial magnetic field leads to an inhomogeneous vortex ring distribution will have an effect on the expected oscillation frequencies.

\section{Discussion}
\label{sec:discussion}

We have shown that it is possible to create and control vortex rings and more complicated vortex structures in three dimensions using the artificial magnetic field around an optical nanofiber.
In addition, and to the best of our knowledge, there is currently no other known method to generate the the structures obtained from non-azimuthally symmetric modes from the linearly and elliptically polarized evanescent fields shown in Figs.~\ref{fig:VortexRings}(b,c) and \ref{fig:HE21_3d}.
We have also shown that the scissors mode of the condensate can be used to detect whether connected vortex structures are present in an elliptic toroidal system.
The structures generated by these fiber-based systems therefore allow one to deterministically design experiments from which one can study complicated superfluid mechanisms, like the kelvin-mode cascade, superfluid turbulence, or reconnection events between superfluid vortex lines.
Being able to stably create these non-trivial vortex configurations may be first step to creating more complex structures like vortex knots in a single-component superfluid BEC system with the optical nanofiber; however, to generate these structures, the magnetic field must have a dependence on $\hat z$, which is not present in our current model.

\section{Acknowledgements}
This work has been supported by the Okinawa Institute of Science and Technology Graduate University and used the computing resources of the Scientific Computing and Data Analysis section. This work has also been supported by JSPS JP17J01488.


\begin{thebibliography}{10}

\bibitem{pethick_smith_2008}
C.~J. Pethick and H. Smith.
\newblock {\em Bose--Einstein Condensation in Dilute Gases}.
\newblock Cambridge University Press, 2 edition, 2008.

\bibitem{madison2000}
K.~W. Madison, F. Chevy, W. Wohlleben, and Jl. Dalibard.
\newblock Vortex formation in a stirred bose-einstein condensate.
\newblock {\em Physical Review Letters}, 84(5):806, 2000.

\bibitem{abo2001}
J.~R. Abo-Shaeer, C. Raman, J.~M. Vogels, and W. Ketterle.
\newblock Observation of vortex lattices in bose-einstein condensates.
\newblock {\em Science}, 292(5516):476--479, 2001.

\bibitem{wacks2014}
D.~H. Wacks, A.~W. Baggaley, and C.~F. Barenghi.
\newblock Large-scale superfluid vortex rings at nonzero temperatures.
\newblock {\em Phys. Rev. B}, 90:224514, Dec 2014.

\bibitem{anderson2001}
B.~P. Anderson, P.~C. Haljan, C.~A. Regal, D.~L. Feder, L.~A. Collins, C.~W. Clark, and E.~A. Cornell.
\newblock Watching dark solitons decay into vortex rings in a bose-einstein
  condensate.
\newblock {\em Phys. Rev. Lett.}, 86:2926--2929, Apr 2001.

\bibitem{bulgac2014}
A. Bulgac, M.~MN. Forbes, M.~M. Kelley, K.~J. Roche, and G. Wlaz\l{}owski.
\newblock Quantized superfluid vortex rings in the unitary fermi gas.
\newblock {\em Phys. Rev. Lett.}, 112:025301, Jan 2014.

\bibitem{ku2016}
M.~J.~H. Ku, B. Mukherjee, T. Yefsah, and M.~W. Zwierlein.
\newblock Cascade of solitonic excitations in a superfluid fermi gas: From
  planar solitons to vortex rings and lines.
\newblock {\em Phys. Rev. Lett.}, 116:045304, Jan 2016.

\bibitem{matthews1999}
M.~R. Matthews, B.~P. Anderson, P.~C. Haljan, D.~S. Hall, C.~E. Wieman, and E.~A. Cornell.
\newblock Vortices in a bose-einstein condensate.
\newblock {\em Phys. Rev. Lett.}, 83:2498--2501, Sep 1999.

\bibitem{yefsah2013}
T. Yefsah, A.~T Sommer, M.~JH Ku, L.~W Cheuk, W. Ji, W.~S Bakr, and M.~W Zwierlein.
\newblock Heavy solitons in a fermionic superfluid.
\newblock {\em Nature}, 499(7459):426, 2013.

\bibitem{abrikosov1957}
A.~A. Abrikosov.
\newblock The magnetic properties of superconducting alloys.
\newblock {\em Journal of Physics and Chemistry of Solids}, 2(3):199--208,
  1957.

\bibitem{tsubota2014}
M. Tsubota.
\newblock Turbulence in quantum fluids.
\newblock {\em Journal of Statistical Mechanics: Theory and Experiment},
  2014(2):P02013, 2014.

\bibitem{barenghi2014}
C.~F. Barenghi, L. Skrbek, and K.~R. Sreenivasan.
\newblock Introduction to quantum turbulence.
\newblock {\em Proceedings of the National Academy of Sciences}, 111(Supplement
  1):4647--4652, 2014.

\bibitem{seo2017}
S.~W Seo, B. Ko, J.~H. Kim, and Y. Shin.
\newblock Observation of vortex-antivortex pairing in decaying 2d turbulence of
  a superfluid gas.
\newblock {\em Scientific reports}, 7(1):4587, 2017.

\bibitem{navon2016}
N. Navon, A.~L Gaunt, R.~P Smith, and Z. Hadzibabic.
\newblock Emergence of a turbulent cascade in a quantum gas.
\newblock {\em Nature}, 539(7627):72, 2016.

\bibitem{ruostekoski2001}
J. Ruostekoski and J.~R Anglin.
\newblock Creating vortex rings and three-dimensional skyrmions in
  bose-einstein condensates.
\newblock {\em Physical review letters}, 86(18):3934, 2001.

\bibitem{shomroni2009}
I. Shomroni, E. Lahoud, S. Levy, and J. Steinhauer.
\newblock Evidence for an oscillating soliton/vortex ring by density
  engineering of a bose--einstein condensate.
\newblock {\em Nature Physics}, 5(3):193, 2009.

\bibitem{ruostekoski2005}
J. Ruostekoski and Z. Dutton.
\newblock Engineering vortex rings and systems for controlled studies of vortex
  interactions in bose-einstein condensates.
\newblock {\em Physical Review A}, 72(6):063626, 2005.

\bibitem{ginsberg2005}
N.~S Ginsberg, J. Brand, and L.~V. Hau.
\newblock Observation of hybrid soliton vortex-ring structures in bose-einstein
  condensates.
\newblock {\em Physical review letters}, 94(4):040403, 2005.

\bibitem{jackson1999}
B. Jackson, J.~F McCann, and C.~S Adams.
\newblock Vortex line and ring dynamics in trapped bose-einstein condensates.
\newblock {\em Physical Review A}, 61(1):013604, 1999.

\bibitem{pinsker2013}
F. Pinsker, N.~G Berloff, and V.~M P{\'e}rez-Garc{\'\i}a.
\newblock Nonlinear quantum piston for the controlled generation of vortex
  rings and soliton trains.
\newblock {\em Physical Review A}, 87(5):053624, 2013.

\bibitem{abad2008}
M. Abad, M. Guilleumas, R. Mayol, and M. Pi.
\newblock Vortex rings in toroidal bose-einstein condensates.
\newblock {\em Laser Physics}, 18(5):648--652, 2008.

\bibitem{dalibard2011}
J. Dalibard, F. Gerbier, G. Juzeli\ifmmode~\bar{u}\else
  \={u}\fi{}nas, and Patrik \"Ohberg.
\newblock Colloquium.
\newblock {\em Rev. Mod. Phys.}, 83:1523--1543, Nov 2011.

\bibitem{mochol2015}
M. Mochol and K. Sacha.
\newblock Artificial magnetic field induced by an evanescent wave.
\newblock {\em Scientific reports}, 5, 2015.

\bibitem{ward2006}
J.~M Ward, D.~G O'Shea, B.~J Shortt, M.~J Morrissey, K. Deasy, and S.~G. Nic~Chormaic.
\newblock Heat-and-pull rig for fiber taper fabrication.
\newblock {\em Review of scientific instruments}, 77(8):083105, 2006.

\bibitem{tong2003}
L. Tong, R.~R Gattass, J.~B Ashcom, S. He, J. Lou, M. Shen, I. Maxwell, and E. Mazur.
\newblock Subwavelength-diameter silica wires for low-loss optical wave
  guiding.
\newblock {\em Nature}, 426(6968):816, 2003.

\bibitem{yariv1976}
A. Yariv.
\newblock {\em Introduction to optical electronics}.
\newblock Holt, Rinehart and Winston, Inc., New York, NY, 1976.

\bibitem{vetsch2010}
E. Vetsch, D. Reitz, G. Sagu{\'e}, R. Schmidt, S.~T. Dawkins, and A. Rauschenbeutel.
\newblock Optical interface created by laser-cooled atoms trapped in the
  evanescent field surrounding an optical nanofiber.
\newblock {\em Physical review letters}, 104(20):203603, 2010.

\bibitem{lacro_te2012}
C. Lacro{\^{u}}te, K,~S. Choi, A. Goban, D.~J Alton, D. Ding, N.~P. Stern, and H.~J. Kimble.
\newblock A state-insensitive, compensated nanofiber trap.
\newblock {\em New Journal of Physics}, 14(2):023056, 2012.

\bibitem{nieddu2016}
T. Nieddu, V. Gokhroo1and, and S. Nic~Chormaic.
\newblock Optical nanofibres and neutral atoms
\newblock {\em Journal of Optics}, 18(5):053001, 2016.

\bibitem{sague2007}
G.~Sagu\'e, E.~Vetsch, W.~Alt, D.~Meschede, and A.~Rauschenbeutel.
\newblock Cold-Atom Physics Using Ultrathin Optical Fibers: Light-Induced Dipole Forces and Surface Interactions.
\newblock {\em Phys. Rev. Lett.} 99(16):163602, 2007

\bibitem{russel2011}
L. Russell, K. Deasy, M.~J. Daly, M.~J. Morrissey, and S. Nic Chormaic.
\newblock Sub-Doppler temperature measurements of laser-cooled atoms using optical nanofibres.
\newblock {\em Measurement Science and Technology} 23(1):015201, 2011

\bibitem{kumar2015}
R. Kumar, V. Gokhroo1, K. Deasy, A. Maimaiti, M. Frawley, C. Phelan, and S. Nic Chormaic
\newblock Interaction of laser-cooled 87Rb atoms with higher order modes of an optical nanofibre
\newblock {\em New Journal of Physics} 17(1):013026, 2015.

\bibitem{kien2004}
F. Le~Kien, V.~I. Balykin, and K. Hakuta.
\newblock Atom trap and waveguide using a two-color evanescent light field
  around a subwavelength-diameter optical fiber.
\newblock {\em Physical Review A}, 70(6):063403, 2004.

\bibitem{phelan2013}
C.~F. Phelan, T. Hennessy, and Th. Busch.
\newblock Shaping the evanescent field of optical nanofibers for cold atom
  trapping.
\newblock {\em Optics Express}, 21(22):27093--27101, 2013.

\bibitem{sachdeva2017}
R. Sachdeva and Th. Busch.
\newblock Creating superfluid vortex rings in artificial magnetic fields.
\newblock {\em Physical Review A}, 95(3):033615, 2017.

\bibitem{cozzini2003}
M. Cozzini, S. Stringari, V. Bretin, P. Rosenbusch, and J. Dalibard.
\newblock Scissors mode of a rotating bose-einstein condensate.
\newblock {\em Physical Review A}, 67(2):021602, 2003.

\bibitem{guery1999}
D. Gu{\'e}ry-Odelin and S. Stringari.
\newblock Scissors mode and superfluidity of a trapped bose-einstein condensed
  gas.
\newblock {\em Physical review letters}, 83(22):4452, 1999.

\bibitem{marago2000}
O.~M. Marago, S.~A. Hopkins, J. Arlt, E. Hodby, G. Hechenblaikner, and C.~J. Foot.
\newblock Observation of the scissors mode and evidence for superfluidity of a
  trapped bose-einstein condensed gas.
\newblock {\em Physical review letters}, 84(10):2056, 2000.

\bibitem{smith2004}
N.~L. Smith, W.~H. Heathcote, J.~M. Krueger, and C.~J. Foot.
\newblock Experimental observation of the tilting mode of an array of vortices
  in a dilute bose-einstein condensate.
\newblock {\em Physical review letters}, 93(8):080406, 2004.

\bibitem{zambelli1998}
F. Zambelli and S. Stringari.
\newblock Quantized Vortices and Collective Oscillations of a Trapped Bose-Einstein Condensate
\newblock {\em Phys. Rev. Lett.}, 81:1754, 1998.

\bibitem{stringari2001}
S. Stringari.
\newblock Superfluid Gyroscope with Cold Atomic Gases
\newblock {\em Phys. Rev. Lett.}, 86:4725, 2001.

\bibitem{minogin2010}
V. Georgievich Minogin and S. Nic~Chormaic.
\newblock Manifestation of the van der waals surface interaction in the
  spontaneous emission of atoms into an optical nanofiber.
\newblock {\em Laser Physics}, 20(1):32--37, 2010.

\bibitem{gpue}
J. Schloss and L.~J. O'Riordan.
\newblock Gpue: Graphics processing unit gross--pitaevskii equation solver.
\newblock {\em Journal of Open Source Software}, 3:1037, 2018.

\end{thebibliography}
\end{document}